\documentclass[10pt,nocopyrightspace,natbib,numbers]{sigplanconf}
\usepackage[table]{xcolor}
\usepackage{graphicx}
\usepackage{amsmath}
\usepackage{hyperref}
\hypersetup{
       colorlinks,
       urlcolor=black,
       citecolor=black,
       filecolor=black,
       linkcolor=black
}

\usepackage{listings}
\usepackage{xspace, ttquot}

\title{
    The Cost of Software-Based Memory Management Without Virtual Memory
}

\authorinfo{Drew Zagieboylo}
           {Department of Computer Science\\ Cornell University}
           {dzag@cs.cornell.edu}
\authorinfo{G. Edward Suh}{Department of Electrical and Computer Engineering\\ Cornell University}{suh@ece.cornell.edu}
\authorinfo{Andrew C. Myers}{Department of Computer Science\\ Cornell University}{andru@cs.cornell.edu}

\definecolor{darkOrange}{rgb}{0.8,0.4,0}
\definecolor{medGreen}{RGB}{112,214,68}
\definecolor{darkGreen}{rgb}{0,0.4,0}
\definecolor{darkBlue}{rgb}{0,0.0,0.5}
\definecolor{medRed}{RGB}{252, 100, 100}
\definecolor{darkRed}{rgb}{0.8,0,0}
\definecolor{salmon}{RGB}{255,160,122}

\begin{document}
\maketitle

\begin{abstract}
Virtual memory has been a standard hardware feature for more
than three decades. At the price of increased hardware complexity, it
has simplified software and promised strong isolation among colocated
processes. In modern computing systems, however, the costs of virtual
memory have increased significantly. With large memory workloads,
virtualized environments, data center computing, and chips
with multiple DMA devices, virtual memory can degrade
performance and increase power usage.
We therefore explore the implications of building applications
and operating systems without relying on hardware support
for address translation. Primarily, we investigate the
implications of removing the abstraction of large contiguous memory
segments.
Our experiments show that the overhead to remove this reliance is
surprisingly small for real programs. We expect
this small overhead to be worth the benefit of reducing
the complexity and energy usage of address translation.
In fact, in some cases, performance can even improve when address
translation is avoided.
\end{abstract}

\section{Introduction}
\label{sec:intro}

Virtual memory is a central abstraction
for modern operating systems (OS), providing processes
with a fictional view of a private, contiguous address space.
Program code references memory
via virtual addresses that are translated by the OS
and hardware to physical memory addresses.
While this abstraction has
provided security, portability, and convenience,
it can be very expensive.
In particular, large-memory applications and virtualized
environments tend to pay the highest performance costs
for address translation~\cite{diyvm,directsegments},
with overheads of up to 94\% reported for microbenchmarks
and 66\% for representative benchmarks.
Graph-processing applications
are hit the hardest, as their poor spatial locality
increases the time spent handling TLB
misses~\cite{vawall}. All these costs are
exacerbated in a virtual machine, since
TLB misses there may require _nested_ page table walks.

Virtual memory also
makes hardware and software designs more complex.
Hardware page table walkers traverse complex
software-managed data structures; modern TLBs
support multiple physical page sizes;
and translation structures must be carefully
managed to enforce security.
To fully utilize these features,
OSes have also accrued more logic and complexity.
For example, the OS must manage page tables and 
ensure that TLBs are synchronized across CPUs and devices.
Additionally, the OS has the difficult job of
turning hardware-level optimizations for translation into
performance gains for software.
Linux originally implemented
transparent huge pages (THP) in 2010, yet the research community continues
to search for practical designs in which THP improves software performance~\cite
{hugepagemgmt, superpages, hawkeye}. While these efforts have continued
to improve the state of the art, they introduce more processes
and heuristics into an already complex kernel.

Embedded systems, with low power and area budgets,
already tend to avoid virtual memory; we posit
that a broader range of systems
can benefit by not translating memory accesses.
Much of the software complexity involved in
virtual memory management is redundant with features
implemented in managed language runtimes, and even in application code.
Therefore, we believe future architectures
can increase performance, simplicity and energy efficiency
by delegating more responsibility
to compilers, language runtimes, and applications.
This approach goes against the grain of the contemporary
trend toward ever more complex hardware optimizations
consuming more power and area.

In this paper, we envision a _physically addressed_
software architecture, without address translation,
in which the OS hands out fixed-size blocks of memory,
and applications allocate data structures across those blocks.
While this is an extreme design, determining the challenges and
costs to software that it implies will be useful in
exploring future hardware support for different memory abstractions.
Our primary contributions to this end are
experiments measuring the performance cost of removing
the abstraction of large contiguous ranges of memory.
Our results suggest that small-footprint applications pay
only a small price from these changes,
while large-footprint applications can benefit from avoiding TLB misses.
We hope that these results incite further research towards supporting
physical addressing without losing the security, convenience, and performance
of virtual memory.

\section{Supporting Virtual-Memory Functionality}
\label{sec:vmperf}

\begin{table}
\caption{Supporting Virtual Memory Features With Physical Addressing}
\label{tbl:vmfunctions}
\begin{tabular}{|p{1.7cm}|p{6.3cm}|}
\hline
VM Function & How To Implement Without Virtual Memory \\
\hline
\hline
Protection & Hardware support for _physical memory protection_ and
OS support for using these features.\\
\hline
Relocation / Migration & Most code is compiled position-independently so that it
can be relocated at run time. For data, managed languages already
include support for relocating objects. CARAT~\cite{carat}
shows how the compiler, runtime and kernel can provide
similar support for unmanaged languages so that
their data can be dynamically relocated. \\
\hline
Swapping & Modern, performance-critical software
is considered non-functional when swapping, so it
is avoided at all cost. However,
language support for relocation directly supports swapping.
Machinery for migrating
objects between memory pages can also
move objects between memory and disk, under application
control. \\
\hline
Contiguity & Without large contiguous
memory regions, language runtimes and programs
represent and access the program stack and large arrays differently.\\
\hline
\end{tabular}
\end{table}
The disadvantages of virtual memory are well known,
and many proposals exist to mitigate them.
Bhattacharjee~\cite{vawall} summarizes
how current virtual memory designs impose performance,
complexity, energy, and area costs, which have become
system-wide bottlenecks for many applications.
In this section, we discuss the implications of changing the memory
allocation abstraction to _physical_ instead of virtual.
Table~\ref{tbl:vmfunctions} summarizes four main features
now provided by virtual memory and how they
can be supported without address translation.
Our conclusion is that
hardware support for physical memory _protection_
may be necessary, but most other features can be implemented using
logic already resident in managed language runtimes. The one
exception is the contiguous-memory abstraction, for which
we propose and evaluate preliminary solutions in Sections~\ref{sec:contiguousmem}
and~\ref{sec:experiment}.

We argue virtual memory can be removed with surprisingly low
performance and complexity costs.
For instance, by adding modest burden
to applications, we can simplify OS memory management.
Recent research on transparently utilizing hardware support
for multi-sized page tables~\cite{hawkeye,hugepagemgmt,superpages}
requires complex OS enhancements to both provide performance improvement
and retain backwards-compatible memory allocation APIs.
However, many applications, such as Memcached,
already manage memory to reduce fragmentation
by exploiting domain knowledge.
Other applications use general-purpose user-space allocators
such as jemalloc. These allocators can
easily be configured to interact with a simple OS memory manager
like the one we describe in Section~\ref{sec:contiguousmem}.

Recently, multiple research projects proposed separating memory
protection from virtual memory to provide stronger security
guarantees, and demonstrated such protection mechanisms can be realized
with low overhead.
For example, capability-based protections, such as CHERI~\cite{cheri},
provide memory protection by including additional meta-data for
each pointer. While capability systems make pointers large, their
overall memory overhead is quite small since they require only
one bit per memory location to distinguish pointers from other data.
Tagged memory such as Hyperflow~\cite{hyperflow} 
and RISC-V's PMP can also provide strong protection by 
attaching security meta-data to each memory chunk. 
Their area overhead is proportional to the granularity of
protection required.
While exact area and power consumption are difficult to quantify, 
we expect that removing (or simplifying)
support for address translation can have a net positive impact since
current translation infrastructure uses as much space as an L1 cache
and up to 15\% of a chip's energy~\cite{vawall}.


A final consideration is flexibility.
Virtual memory is baked into the ISA,
and so are performance-influencing parameters like page size.
Modern instruction sets provide only a few possible page sizes.
For example, x86\_64 only supports 4\,KB, 2\,MB or 1\,GB pages;
ARMv8 only supports 4\,KB, 16\,KB or 64\,KB.
This illustrates one of the problems with an otherwise simple
solution to some virtual memory woes: increasing a system's
base page size such that _all pages_ are huge pages.
To achieve the best TLB reach with the least wasted memory,
the OS would ideally choose a page size parameterized on available resources and workload;
however, this is not possible in general with
the limited options provided by hardware.
Physical addressing gives the OS more choice
as technology, memory resources, and workloads vary over time.

\section{Contiguous Memory}
\label{sec:contiguousmem}

We imagine that realistic physically addressed systems could utilize
techniques from prior work~\cite{hawkeye,hugepagemgmt,superpages,directsegments}
to implement flexible partitioning of memory, and a rich API
for reserving contiguous regions of various sizes.
Nevertheless, for the rest of this paper, we describe a
general-purpose OS with a more straightforward
memory allocation strategy: segment memory into
fixed-size blocks as the minimum allocation unit.
In our experiments (Section~\ref{sec:experiment}),
performance was mostly insensitive to the choice of block size
and we report results based on 32\,KB blocks.
In a real deployment, this number would likely be of similar
magnitude (somewhere between small and huge pages)
but not necessarily the same.
While simplistic, this strategy provides a useful tool for exploring
how physically addressed applications might interact
with a constrained memory manager.
Since this OS has less control over external fragmentation, it cannot
provide the conventional expectation that 
arbitrarily large memory requests are satisfied as
long as there is enough unallocated memory.
We investigate the performance cost of modifying software to avoid
large allocations. Our measurements on a variety of benchmarks
suggest that this cost is surprisingly small.
Furthermore, unlike traditional address translation,
this cost is only paid on accesses to ``contiguous''
structures; all other memory accesses incur no overhead.
Modifications are needed to address two key
uses of contiguous memory in programming languages:
the program stack and large arrays.

\subsection{Dynamically Allocated Stack}
\label{sec:stacksplitting}

Stack-relative addressing normally requires contiguous memory locations,
but only within a given stack frame. Therefore, as
long as every stack frame fits within the block size,
code only needs to be modified to dynamically allocate memory
when the current stack block runs out of space.
This modification adds some overhead to each function call (about three x86 instructions)
to ensure the current stack block has enough space.
In the rare case that it doesn't, a new frame is allocated, non-register arguments are
copied from the old stack to the new frame, and the stack pointer is
adjusted; at function exit, all of this work is cleaned up.
By carefully managing the return address register on function entry,
the cleanup code can be skipped when a new block is not allocated.

An existing option of the gcc compiler,
_stack splitting_, already implements this functionality.
It was originally developed to reduce memory usage in highly
parallel programs, but it fulfills our purpose as well.
The only modification we make to gcc's implementation is to 
force new stack memory requests to be equal to the block size.
Although we could allocate perfectly sized frames for each function,
it would increase the number of calls made to the allocator.
Amortizing these calls by using larger allocations usually improves performance
but may create internal fragmentation within the stack memory.

\subsection{Large Array Allocations}
\label{sec:arrayallocations}

We must also consider how applications will
create arrays on the heap, without the
assumption that allocation (e.g., via "malloc") can
return arbitrarily large contiguous regions.
A straightforward approach is to replace arrays
with trees, where intermediate nodes in the tree hold pointers
to other nodes and only the leaves actually store data.
In a sense, hardware-supported page tables implement a similar
data structure; we investigate replacing it with
a purely software version only used for large arrays.

\begin{figure}
       \includegraphics[width=\columnwidth]{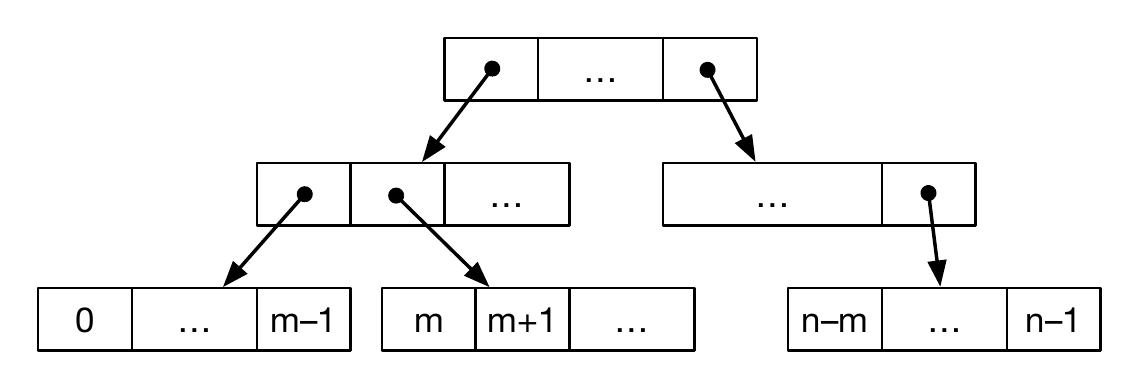}
       \vspace{-2em}
       \caption{Memory layout for an array of length "n" represented as a tree.  Each sequence of boxes represents
        a contiguous allocation of size "m".
A tree stores meta-data about its depth. Data is stored exclusively on leaf nodes; intermediate nodes store
indirection pointers.}
        \vspace{-1em}
        \label{fig:layout}
\end{figure}

Our data structure is an implementation of ``arrays as trees'',
a discontiguous array design described by Siebert~\cite{treearray}.
Figure~\ref{fig:layout} provides a graphical representation of how
data is partitioned across multiple memory allocations in the tree.
Trees are built with constant-sized allocations, independent of
the data size. Supporting a larger number of elements
requires a deeper tree, but this is still bounded by a relatively small
number\footnote{
With the 32\,KB node size used in this work,
3-level and 4-level trees can address
about 536\,GB and 2\,PB of data, respectively.
}.

\begin{figure}
\begin{lstlisting}[basicstyle=\scriptsize\tt]
struct ArrayIterator<int> {
  size_t nextPtr, lastPtr; //pointers to cached elements
  size_t lastIdx; // index of data pointed to by lastPtr
}
ArrayIterator<int> it; Tree<int> tree;
int next() {
  if (it.nextPtr > it.lastPtr) {
     DataPage dp = tree.getDataPage(it.lastIdx + 1);
     it.nextPtr = dp.firstPtr; it.lastPtr = dp.lastPtr;
     it.lastIdx = dp.size + it.lastIdx;
  }
  return *it.nextPtr++;
}
\end{lstlisting}
\caption{Iterator "next()" implementation}
\label{fig:iterexample}
\end{figure}

In general, accessing an element in the tree
requires traversing a path from the root to a leaf and making
one memory access for each layer; however,
software optimizations can reduce this overhead significantly for common cases.
For instance, when iterating sequentially,
software can cache a pointer to the most recently accessed
element. As long as it is part of the same
allocation, software only needs to increment this pointer and make
a single memory access to retrieve its data.
A full tree traversal happens only when iterating
past the last element in a given allocation.
This optimization can be captured abstractly
through an "Iterator" interface,
so that standard iteration constructs in modern programming
languages can use this feature transparently.
Figure~\ref{fig:iterexample} is a pseudocode
implementation of "next()" method of "Iterator", which retrieves
the next tree element. Inlined, this method offers further
optimization opportunities.

\section{Experimental Results}
\label{sec:experiment}

We ran all experiments on a computer with
16 Intel i7-7700 CPUs clocked at 3.60\,GHz, 32\,KB of L1 instruction and data caches,
and 128\,GB of physical memory, running Ubuntu 18.04.
All reported measurements are the average of ten runs, but with
standard configurations to reduce variability
(such as disabling ASLR and powersave mode),
all sample standard deviations
were less than 0.1\% of the mean.

\subsection{Dynamic Stack Allocation Overhead}
\label{subsec:stack}

We consider standard benchmarks to evaluate realistic impact,
and also one microbenchmark designed to study a pessimistic case.
Our standard benchmarks are most of the SPECInt2017 and PARSEC
suites. We omit the ``exchange'' SPEC benchmark because
it is written in FORTRAN, and also ``perlbench'' and
``gcc'' since they crash with gcc's implementation of stack splitting.
Our microbenchmark is designed to amplify the performance cost of stack splitting
beyond what would be seen in most programs;
it is a recursive Fibonacci program
that allows measuring the overhead of 
checking available stack space
in function-call-bound code.

\begin{figure}
        \centering
        \includegraphics[width=\columnwidth]{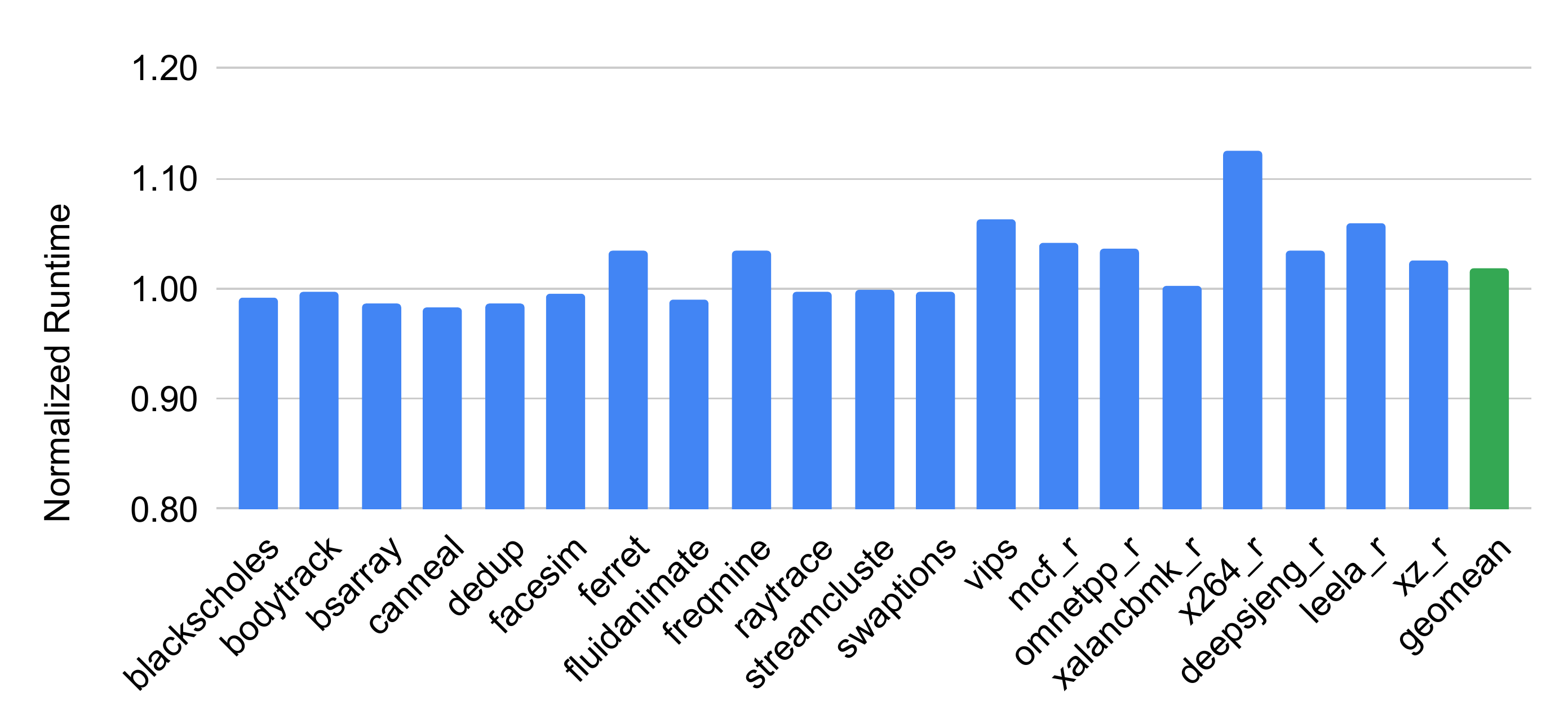}
        \vspace{-2em}
        \caption{Split-stack overhead on PARSEC and SPECInt2017.}
        \label{fig:splitstack}
\end{figure}

Figure \ref{fig:splitstack} shows the run time of split-stack compiled programs
normalized to the default gcc-compiled run times.
The average run-time increase was only 2\%.
The variability seen depends upon the frequency with which the
programs make function calls.
In most cases the performance changed by less than 1\%,
which we believe is essentially noise---it is less than the impact
of changing stack alignment~\cite{noharm}.
We did modify one benchmark (``ferret'') to change very large stack-based allocations
to heap-based allocations in both the baseline and split-stack executions.
These results validate our hypothesis that stack splitting
is unlikely to add significant overhead.
Even the Fibonacci microbenchmark showed only a 15\% slowdown, which seems
acceptable considering its pathological nature.

\subsection{Tree Access Overhead}
\label{subsec:arrays}

\begin{table}
  \caption{
           Ratios of run times for simulated physical-memory
           tree-based implementations vs.\@ virtual-memory implementations.
           Tree-based implementations of arrays are compared against
           traditional contiguous arrays; 4\,KB arrays
           fit into depth-1 trees, 4\,MB into depth-2 and all others in depth-3.
           For each of the benchmarks we provide both a naive implementation
           and a corresponding iterator optimized version. Cells which are 10\%
           slower or faster than the baseline are colored for clarity.
  }
\begin{center}
  \label{fig:micro}
  \begin{tabular}{|m{1.5cm}|m{0.5cm}|m{0.5cm}|m{0.5cm}|m{0.5cm}|m{0.6cm}|m{0.6cm}|m{0.6cm}|}
   \hline
   Benchmark & 4KB & 4MB & 4GB & 8GB & 16GB & 32GB & 64GB \\
   \hline
   \hline
   Linear Scan: Naive  & \cellcolor{medRed} 1.36 & \cellcolor{medRed} 2.97 & \cellcolor{medRed} 3.34 & \cellcolor{medRed} 3.37 & \cellcolor{medRed} 3.37 & \cellcolor{medRed} 3.37 & \cellcolor{medRed} 3.37  \\
   \hline
   Linear Scan: Iter   & 1.00 & 1.02 & 0.99 &  0.99 &  0.99 & 0.99 &  0.99  \\
   \hline
   Strided Scan: Naive & \cellcolor{medRed} 1.71 & \cellcolor{medGreen} 0.72 & \cellcolor{medRed} 1.28 & \cellcolor{medRed} 1.26 & 1.08 & 1.04 & 1.06  \\
   \hline
   Strided Scan: Iter  &  \cellcolor{medRed} 2.47 & \cellcolor{medGreen} 0.57 & 1.02 & \cellcolor{medGreen} 0.89 & \cellcolor{medGreen} 0.86 & \cellcolor{medGreen} 0.86 & \cellcolor{medGreen}0.86  \\
   \hline
  \end{tabular}
  \end{center}
\end{table}

\begin{figure}
        \includegraphics[width=\columnwidth]{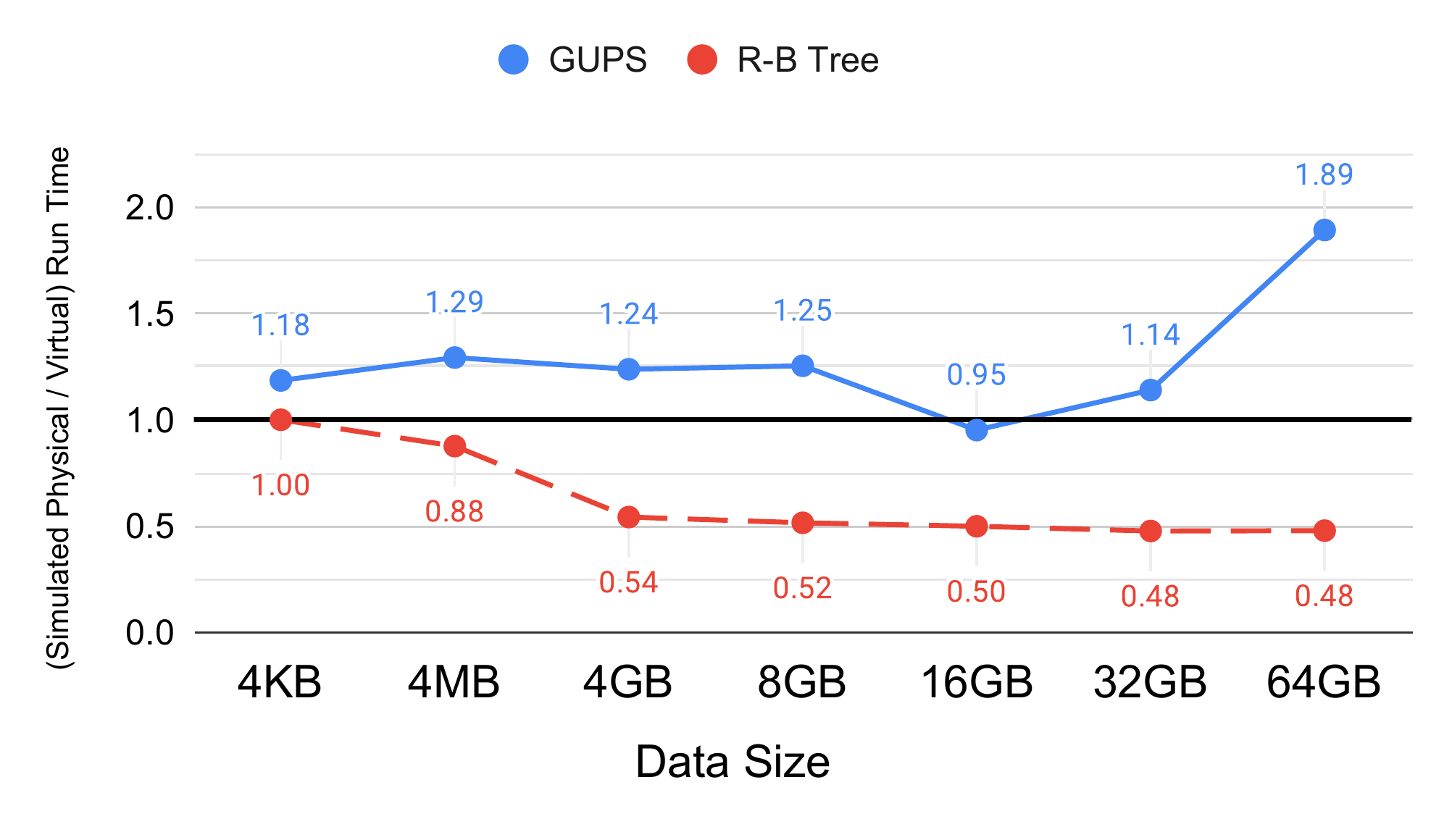}
        \caption{Ratios of run times for two large data structures:
        GUPS and red--black trees. For GUPS, a simulated physical-memory
        tree-based implementation of GUPS is compared to the virtual-memory implementation.
        For red--black trees, the same implementation is used for
        both physical and virtual memory.
        In both cases, the results suggest that physical addressing
        offers better performance for large data structures.
        }
        \label{fig:red-black-results}
        \label{fig:gups-results}
\end{figure}

\begin{figure}
        \centering
        \includegraphics[width=\columnwidth]{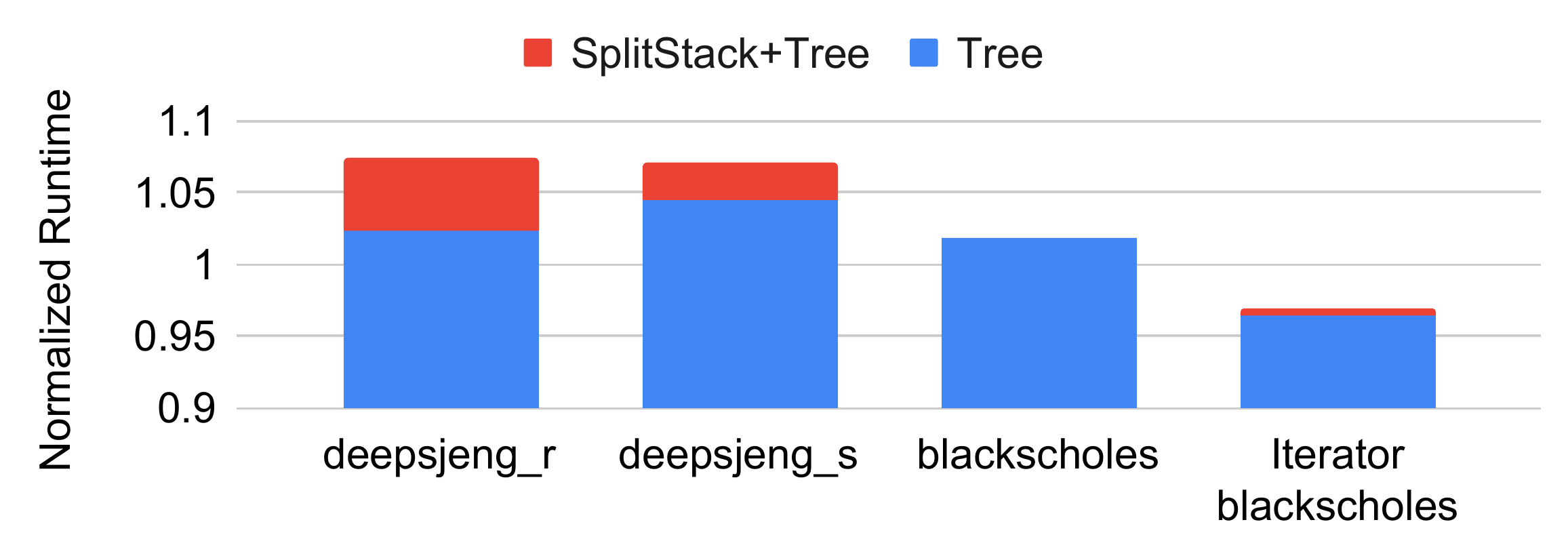}
        \caption{Overhead of software-based contiguous memory on selected SPEC and PARSEC benchmarks.}
        \label{fig:bench}
\end{figure}

To test the impact of replacing large arrays with trees,
we evaluated both microbenchmarks and some standard benchmarks for
runtime overhead. In order to more accurately simulate the physically addressed system
that we imagine, we execute the tree-based implementations of our
microbenchmarks using 1\,GB huge pages to reduce the TLB miss rate to 0,
in most cases.
For the baseline contiguous array implementations, we _did not_ use huge pages.
For standard benchmarks, we used 4\,KB pages for both arrays and trees;
TLB miss rates were always close to 0 regardless of page size.

Our microbenchmarks exhibit various levels of spatial
locality by: (1) iterating over every element; (2) accessing every 1024\textsuperscript{th}
element (i.e., 4\,KB apart); and (3) accessing pseudorandomly (in GUPS, an HPC benchmark).
Lastly, we include a red--black tree benchmark which does not use an array implementation
in either experiment to illustrate the potential speedup of removing virtual memory when
contiguity is not necessary. It creates a red--black tree by inserting random elements and then executes
an in-order traversal that accesses memory locations with low locality.

We evaluate two standard benchmarks which exhibit good and bad spatial locality, respectively:
blackscholes from PARSEC; and deepsjeng from SPECInt2017.
The former scans through several large arrays while executing 
floating point computations on each element; the latter allocates a single
large array as a hashtable and accesses it less predictably.

We measured the average element access time for each
microbenchmark; Table~\ref{fig:micro} compares how
the tree and contiguous array implementations perform
on the two scan benchmarks, both with and without
software optimizations for tree iteration.
For depth 1 trees (4\,KB), we might expect
_no_ overhead, since they require no more memory accesses
than arrays do. However, our implementation checks the
depth of the tree before accessing data, which adds branch instructions
on every access. Similarly, some of our optimizations cause unnecessary
overhead on very small trees. If we statically remove these operations, depth-1 trees
have identical performance to arrays in all benchmarks.
Compilers and language runtimes should be able to eliminate many tree-depth checks
via static analysis and code specialization~\cite{JGS93}.

For deeper trees, we saw an expected jump in overhead caused by the increased
number of memory accesses, in both of the unoptimized tree implementations.
The "4MB" strided data point is the outlier, where trees outperform contiguous
arrays substantially. This occurs because the contiguous arrays see the same overhead
caused by TLB misses on all experiments (except for "4KB") but the "4MB" trees
require fewer memory accesses than all larger data points.
In the linear scan, the arrays suffered almost no TLB
misses, obviating the main advantage of trees.
We hypothesize that translation hardware is
optimized to make this case fast. Nevertheless, in this benchmark,
trees added no overhead when using the "Iterator" optimization.
We observed a similar phenomenon in the strided experiment, where
arrays suffered extremely high TLB miss rates (above 90\%) but didn't
experience as much slowdown as we expected and thus outperformed
the naive tree implementation.
Likely, hardware optimizations (such as page table walk caches and prefetchers)
reduced the time to handle each TLB miss, mitigating some of
the performance problems caused by the strided scan with contiguous arrays.
Again, by using the "Iterator" optimization, trees running on physical memory
even managed to outperform the original array implementation.

In Figure~\ref{fig:gups-results}, we make the same comparisons but
on the GUPS and red--black tree benchmarks. These benchmarks have random
access patterns that should both cause significant TLB misses
and make hardware translation optimizations less effective.
As expected, using a tree implementation does cause less slowdown
on GUPS; trees even _outperform_ arrays for the 16\,GB GUPS dataset,
so physical addressing should perform better at that size or
larger. Our red--black tree benchmark, which used the same
non-contiguous implementation in both experiments,
saw up to a 50\% reduction in run time when running without virtual memory.
These results indicate the unsurprising,
but hopeful fact that removing all time spent
handling TLB misses can greatly improve performance even for
workloads of relatively modest size.

Figure~\ref{fig:bench} summarizes the performance overhead induced
by removing the contiguous memory abstraction on blackscholes and deepsjeng.
Allocations by blackscholes totaled 600\,MB of memory,
deepsjeng\_r uses 700\,MB, and deepsjeng\_s uses 7\,GB.
In all cases, replacing large arrays with trees degraded performance by less than 3\%;
performance even improved slightly for blackscholes implemented with "Iterator"s.
Even with stack splitting, total overhead is under 10\%.

\subsection{Experimental Limitations}

Note that in our experiments, trees perform worse than arrays
on the 32\,GB and 64\,GB datasets for GUPS and strided accesses.
This is an artifact of our experimental setup; beyond 16\,GB, huge pages don't faithfully simulate
physical-memory performance because they start taking TLB misses too~\cite{directsegments}.
Essentially, in these datapoints we're seeing the software overhead of trees but
none of the benefits that physical addressing would convey.
While we would have preferred to run these experiments on ``bare metal'' hardware
without virtual addressing, it is impractical to achieve such a setup
while running normal software with huge datasets.
Based on the performance trends we did observe, we believe
that trees would start to outperform arrays for these two benchmarks
on huge datasets with _actual_ physical memory.
Tree depth can effectively be held constant, but the cost in TLB misses
is only going to rise as datasets get even larger.

\subsection{Discussion}
Our evaluation measures overheads arising because of the loss
of the contiguous memory abstraction. Our selected SPEC and PARSEC
benchmarks indicate that CPU-bound workloads see little overhead
from trees and split stacks. The overhead from implementing medium-sized
arrays as trees can be noticeable, but should be smaller or
even negative in large-memory applications.
Further, many applications (e.g., graph processing)
act more like the red--black tree
benchmark; our results suggest they will run significantly faster
in physical memory.

Additionally, the performance gap that does remain in some
of our experiments is clearly a result of differing performance
optimizations. Modern address translation hardware is complex
and heavily optimized, especially for common access patterns.
In addition to TLBs, page table walk (PTW) caches store intermediate
page table information to speed up future page table traversals.
Prefetching also helps to hide TLB miss latency when access patterns are predictable.
However, we can implement similar optimizations in software,
achieve similar results, and avoid bloating hardware with
wasteful complexity. The Linear Scan:Naive and Linear Scan:Iterator
experiments exemplify this perfectly. Our Iterator optimization essentially
implements a PTW cache in software, so that we rarely traverse the
whole tree; applying this optimization to the Linear Scan application completely
removes the performance gap between traditional virtual memory and our simulated physical system.
In general, it is better for applications to target similar optimizations
whenever they are actually needed, rather than expending circuitry
on hardware to _predict_ which optimizations might be beneficial at the moment.
Certainly, there are are inherently unpredictable programs (like GUPS) where no static optimization
can help. Peformance of those cases could
benefit from hardware acceleration of tree traversals, perhaps using
some simplified subset of current virtual memory optimizations. Nevertheless,
making that functionality an optional accelertor rather than an obligate
step on the critical path to memory could offer the best of both optimization schemes.

\section{Related Work}

The many efforts to address the ``address translation wall''~\cite{vawall}
have largely been directed at opaquely improving translation hardware.
More relevant to this paper are efforts to modify or remove traditional address
translation, or those that offer hardware support to enable such a transition.

CARAT~\cite{carat} consists of a compiler, runtime and kernel module,
which together provide memory protection and the ability
to relocate memory despite the program using physical addresses.
In this way, CARAT preserves the virtual memory abstraction for application
software but removes the dependence on hardware support for address
translation and memory protection (within the application).
CARAT inserts runtime allocation tracking and guards, and uses
a sohpisticated program dependency analysis to support these
features with minimal overheads.
Our work suggests a more radical change, to entirely remove
software dependence on hardware translation. These techniques are
certainly compatible and the most performant future systems
will likely make use of both. For instance, ``arrays as trees'' could
ameliorate an existing CARAT limitation that
arises from the difference in the sizes of application allocations
versus those of the underlying allocator.

Basu et al.~\cite{directsegments} use direct segments to enable efficient contiguous
memory in the face of address translation.
These support efficient translation of large, contiguous memory regions
allocated by the OS via base and bound registers. While effectively eliminating
translation overhead for certain accesses, this scheme is still essentially static in
nature, relying on traditional virtual memory as the default system behavior.

Alam et al.~\cite{diyvm} propose a ``Do-It-Yourself''
translation mechanism that allows application software to
provide its own address translation function. The hardware then checks
that the result of translation is a legal, accessible physical frame.
Their work provides more flexibility than traditional virtual memory
abstractions, but 
still fundamentally relies on address translation and thus is still
likely to perform worse than direct physical memory.

Many other hardware-enabled protection schemes do not rely on
address translation and thus are good candidates to support protection
in the face of physical addressing. Mondrian memory protection~\cite{mondrian}
efficiently provides fine-grained partitioning and sharing of memory and would enable
flexible OS memory allocation. Other results~\cite{efficient,cheri,timber}
indicate that physically tagged memory can provide efficient
protection without relying on address translation; these approaches
could support a variety of software memory-management abstractions.

\section{Conclusion}
\label{sec:conclusions}

Virtualization, large memory footprints,
and coherency across multiple cores and devices
all make address translation a growing performance bottleneck
and source of complexity.
We have explored an alternative approach:
physical addressing while delegating responsibility to software.
This direction has been insufficiently examined by the architecture, OS,
and programming language communities. Our experimental results show that
the software overheads added by replacing the contiguous memory abstraction are low
and that physical addressing is potentially a fruitful area for further research.
While we have examined a narrow set of changes
to applications and compilers, there are many more opportunities
to optimize memory management and access at all layers of the programming stack.
Additionally, removing hardware-based address translation would
enable simpler, more power-efficient circuits whose functionality
and usage is driven by application needs rather than constrained by rigid
design choices.

\section*{Acknowledgments}
This work was supported by the Department of Defense (DoD)
through the National Defense Science \& Engineering Graduate Fellowship (NDSEG) Program.

\bibliographystyle{abbrvnat}
\bibliography{novm,bibtex/pm-master}

\end{document}